\documentclass[preprint]{aastex631}

\newcommand{\cgro}{\mbox{\textit{CGRO}-BATSE~}}

\newcommand{\fermi}{\mbox{\textit{Fermi}-GBM~}}

\newcommand{\bB}{{\boldsymbol{B}}}

\usepackage{comment}
\usepackage{amsmath, soul}
\usepackage{multirow}

\graphicspath{{./}{figures/}}

\shorttitle{Magnetar Burst Clusters}
\shortauthors{Keskin et al.}

\begin{document}

 \title{Investigating the Recursive Short X-ray Burst Behavior of Magnetars Through Crustal Interactions}

\correspondingauthor{\"Ozge Keskin}
\email{ozgekeskin@sabanciuniv.edu}

\author[0000-0001-9711-4343]{\"Ozge Keskin}
\affiliation{Sabanc\i~University, Faculty of Engineering and Natural Sciences, \.Istanbul 34956 T\"urkiye}

\author{Samuel K. Lander}
\affiliation{School of Engineering, Mathematics and Physics, University of East Anglia, Norwich, NR4 7TJ, U.K. }

\author[0000-0002-5274-6790]{Ersin G\"o\u{g}\"u\c{s}}
\affiliation{Sabanc\i~University, Faculty of Engineering and Natural Sciences, \.Istanbul 34956 T\"urkiye}


\begin{abstract}

Energetic bursts from strongly magnetized neutron stars, known as magnetars, are typically detected in clusters. Once an active episode begins, anywhere from a few to thousands of hard X-ray bursts can occur over durations ranging from days to months. The temporal clustering of these recurrent bursts during an active episode suggests an underlying mechanism that triggers multiple bursts in rapid succession. These burst clusters are likely crucial for understanding the processes driving magnetar activity. In this study, we investigate the repetitive short X-ray burst behavior of magnetars through crustal interactions, employing the cellular automaton model for the magnetar crust proposed by \citet{Lander23}. Our simulations, based on physically motivated criteria, successfully reproduce burst clustering. Additionally, the durations and energetics of active episodes in our simulations agree well with observational data. We discuss the potential physical mechanisms underlying burst clusters observed in numerous magnetars, as well as the reactivations of an individual magnetar. 

\end{abstract}

\keywords{Neutron Stars (1108), Magnetars (992), X-ray bursts (1814)}


\section{Introduction} \label{sec:intro}

Magnetars are isolated, slowly rotating (2-12 s), young ($\sim 10^3$ yr), and strongly magnetized (B $\sim10^{14}-10^{15}$ G) neutron stars \citep{DT92, Kouveliotou99, Olausen2014, Kaspi2017}. Their extremely strong magnetic fields are thought to play a pivotal role in their emission of 
highly luminous transient events in hard X-rays/soft $\gamma$-rays; such as short bursts, outbursts, burst storms/forests, and intermediate \& giant flares,
as well as their persistent emission. The most frequently observed high-energy magnetar phenomena are the short bursts with durations ranging from a few milliseconds to a few seconds and energies between $\sim10^{38}$ and $10^{41}$ erg \citep{EG01}. 
Both time-integrated and time-resolved spectral studies so far have shown that short bursts can be modeled nearly equally well with a thermal model, such as the sum of two blackbodies, or a nonthermal model, such as a power law with a high-energy exponential cutoff \citep{Israel08, van12, Younes14, Lin2016, Lin2020, Keskin2024}. 
The rarely observed but the most energetic magnetar activity is a giant flare, during which magnetars release energies of $\sim10^{44-46}$ erg in a few hundred seconds \citep{Hurley99, Palmer05}. Between these two types of high-energy transients in terms of duration and energy, intermediate flares last from a few seconds to tens of seconds with energies of $\sim10^{42}$ erg \citep{Alaa01, CK01}. Additionally, magnetars such as SGR 1900+14 \citep{Israel08}, SGR 1550-5418 \citep{YK2010,van12}, and SGR 1935+2154 \citep{YK21} have shown intense high-energy burst storms/forests.

Typical magnetar bursts are generally associated with local or global yielding of the solid neutron star crust through the release of the built-up elastic stress in the crust due to the internal magnetic field evolution \citep{TD95, TD01}. Another possible mechanism to generate magnetar bursts is the rapid untwisting of the twisted external magnetic field lines \citep[magnetic reconnection,][]{TD95, lyu03}. The understanding of magnetar activities is still fragmented and incomplete. This is due to the fact that both the “triggering” mechanism of a magnetar event and how the triggered system generates outgoing and evolving radiation within the magnetosphere are unclear.

Considering the broad variety of magnetar activities, it is possible that different types of events have distinct physical origins. These make understanding the underlying physics challenging and hinder the construction of a unified physical model to explain a diverse range of activities. Motivated by this issue, \cite{Lander23} recently presented a cellular-automaton model that aims to explain the wide range of energies of magnetar activity through the yielding of the star's crust. Building and release of elastic stress from the crust leads to the braiding of coronal loops, which ultimately induce reconnection events and produce the energy releases we observe. While the local nature of coronal loops allows for the possibility of high-energy short bursts, the model also includes a criterion that causes more widespread yielding -- leading to giant flares -- and quiescent periods which are typically decades long.

One notable feature of magnetar bursts is that they often do not come in isolation: in many cases, we observe clusters of bursts, often separated in time by a second or less \citep[see e.g.,][]{van12, Lin2020}. These burst clusters may be a key to understanding the underlying processes driving magnetar activity, since such temporal concentrations of bursts are highly unlikely to be coincidental, but rather indicate a mechanism that causes the triggering of several bursts in quick succession. In this paper, we investigate the recursive short X-ray burst behavior of magnetars through crustal interactions by utilizing the magnetar simulation of \cite{Lander23}. We discuss the possible underlying physical scenarios responsible for magnetar burst clusters as seen in many magnetar observations. The flow of this paper is as follows: In Section \ref{sec:B_evol}, we summarize crustal magnetic field evolution, in Section \ref{sec:sam_model}, we provide a brief summary of the relevant details of the cellular automaton model, and in Section \ref{sec:small_bursts}, we provide a physical picture of magnetar burst clusters. In Section \ref{sec:model}, we explain the model improvements and tests to study burst clusters. Finally, the results are presented in Section \ref{sec:res} and discussed in Section \ref{sec:dis}.


\subsection{Crustal Magnetic Field Evolution} \label{sec:B_evol}

Much work modeling magnetars has focused on the evolution of the magnetic field within the star's crust. In particular, by assuming the crust is rigid -- so that ions are static, and the electrons are the only mobile charged particle -- `electron MHD' may be derived. The crustal magnetic field then evolves according to the equation \citep{gold_reis92}:
\begin{equation}\label{eMHD}
    \frac{\partial \bB}{\partial t}=-\frac{1}{4\pi}\nabla\times\left(
    \frac{c}{\rho_e}(\nabla\times\bB)\times\bB
    -\frac{c^2}{\sigma}\nabla\times\bB
    \right)
\end{equation}
where $\rho_e$ is the charge density and $\sigma$ the electrical conductivity. Equation \eqref{eMHD} represents an interplay between Hall drift (the first term on the right), causing the field to evolve towards more intense high-multipole structures, and Ohmic decay (the second term on the right), which dissipates magnetic flux. Generally speaking, Ohmic decay dominates for weaker fields. Beginning from the early axisymmetric simulations, the study of electron MHD has advanced greatly, including full 3D nonaxisymmetric simulations, and better handling of the numerically challenging Hall term \citep{Wood_hollerbach2015, Gourgouliatos2016, Degrandis2020, Igoshev2021, Dehman2023, Ascenzi2024}. Furthermore, because Ohmic decay causes heating, and the conductivity is temperature-dependent, the full electron-MHD system becomes a coupled magnetothermal evolution; see \citet{pons_vig_19} for a review of this progress. The attractive feature of these simulations is that whilst they are numerically challenging, they involve only a small number of well-understood transport properties of the crust.

For a subset of the possible initial conditions and boundary conditions, magnetothermal evolutions lead to the development of intense patches of magnetic field on kyr timescales, with Maxwell stress components that are comparable to the crust's elastic yield stress $\tau_{el}$ \citep{Chugunov-Horowitz2010}. This provides persuasive evidence that crustal field evolution drives crustal failure and so the observed magnetar bursting activity. By resetting the local magnetic field when stresses build beyond some critical value, \cite{Perna-Pons-2011} arrived at simulated event rates and waiting times between bursts (see also \cite{dehman20}), and identified factors that likely result in more or less frequently-bursting magnetars.

However, one should be cautious in simulating crustal failure with electron MHD, which itself is built on the assumption that the crust does \emph{not} fail. To be self-consistent, the evolution beyond the failure point should also include a mechanism that describes the motion of the crust, with some velocity $\boldsymbol{v}$, under suprayield stresses. This in turn causes advection of the field of the form $\nabla\times(\boldsymbol{v\times B})$, which should be added to the right-hand side of Equation \eqref{eMHD}. We are, however, interested in the case when high stresses can develop and be released through plastic flow, and this process requires Ohmic decay to be sub-dominant. Given the high temperatures of typical young magnetars, this corresponds to field strengths of the order $10^{14}\,\mathrm{G}$ \citep{cumming04}. We explore the implications of the interplay of Hall drift and Ohmic decay for our burst model in two follow-up studies (Keskin et al, in prep.; Lander et al, in prep). For now, however, we neglect Ohmic decay for simplicity, so that the resulting field evolution takes the form \citep{L16}:
\begin{equation}\label{magnetoplastic}
    \frac{\partial \bB}{\partial t}=-\frac{1}{4\pi}\nabla\times\left(
    \frac{c}{\rho_e}(\nabla\times\bB)\times\bB
    \right)
    +\nabla\times(\boldsymbol{v\times B}),
\end{equation}
where $\boldsymbol{v}=\boldsymbol{0}$ below the yield stress.
Unlike electron MHD, this is no longer a system with a limited set of well-constrained parameters. The main line of work studying the evolution of Equation \eqref{magnetoplastic} to date \citep{Lander2019, Gourgouliatos2021, kojima21} has calculated a velocity $\boldsymbol{v}$ under some simplifying but plausible assumptions that reduce it to the dynamics of a viscous and incompressible fluid \citep{L16}. Even this simplified system, however, has uncertainties related to the unknown material properties of the crust at high stress and the fact that there is no universal theory of viscoplasticity that can be readily applied.


\subsection{Cellular Automaton Magnetar Simulation} \label{sec:sam_model}

As discussed above, electron-MHD simulations show that stresses can quickly grow beyond the yield stress \citep{Perna-Pons-2011, Lander2019, gour23, dehman20}, after which the simulations are physically unreliable, as their inbuilt assumption of a static crust is violated. Furthermore, the standard paradigm for magnetar activity links their sudden bursts with failures of the crust. Crustal failure presents one particular conceptual challenge: whilst first-principles simulations indicate that the ion lattice fails collectively on a microscopic scale, this cannot extend to the global scale (otherwise, a very localized supra-yield stress could cause the entire crust to fail). With quantitative modeling very far away from being able to attack this problem, one alternative is to take a qualitative description of the evolution of crustal stress locally, and formulate a system where local regions can communicate with one another to produce larger-scale and non-deterministic behavior. This was the goal of an earlier cellular automaton model of the crust \citep{Lander23}, whose main features we review next, including where specific choices may reduce the generality of the model.

To simulate the above-mentioned approach, \cite{Lander23} divided the crust into an array of semi-autonomous cells. Within each cell, physical quantities are assumed constant and fixed to their value at the base of the cell. In both the original paper and the present one, the stellar structure is calculated from the TOV equation using the SLy4 equation of state, the yield stress from the formula of \cite{Chugunov-Horowitz2010}, and the viscosity of the crustal matter in its plastic phase $\nu$ with the same profile as that of $\tau_{el}$, together with the temperature-dependent prefactor suggested by \cite{Lander23}. Guided by the results of 3D electron-MHD simulations \citep{Gourgouliatos2016}, which show the development of locally intense patches of magnetic field for young magnetars, whose angular extent has a characteristic diameter  $\sim 0.5-2\,\mathrm{km}$, we fix the surface area of each cell to 1 km$^2$.

Because the outer crust cannot sustain high stress and has a lower melting temperature, only the inner crust is modeled. Equally, however, across the whole inner crust of thickness $\sim 0.5\,\mathrm{km}$, the yield stress changes by a factor of a thousand, meaning it is physically unlikely that any small failure would extend down to the crust-core boundary. For this reason, the cell depth is chosen to be smaller than this, 0.2 km, such that the yield stress variation is less than an order of magnitude. The density within a cell ranges from $4.0\times 10^{11}\,\mathrm{g}\,\mathrm{cm}^{-3}$ at the top, down to $1.5\times 10^{13}\,\mathrm{g}\,\mathrm{cm}^{-3}$ at the base. Whilst the precise choice of depth is arbitrary, it is a plausible value over which collective failure of an entire cell could be expected (as required for a cellular-automaton model).

The magnetoplastic evolution in Equation \eqref{magnetoplastic} may be reduced, in an approximate way, to an equation for a scalar stress $\tau\equiv B^2/8\pi$ that is assumed to be sourced by the magnetic field alone (see \cite{Lander23} for derivation):
\begin{equation}\label{dtau_dt}
    \frac{\partial \tau}{\partial t}=\frac{c \tau^{3/2}}{\pi^{1/2} \rho_e L^2} -  \frac{2 \tau \left(\tau - \tau_{el}\right)}{\nu}
\end{equation}
where the first term on the right-hand side represents the increase of elastic stress due to the Hall drift, while the second term reflects the stress release due to the plastic flow. 
Because plastic flow is inactive below the yield stress but dominant over Hall drift above the yield stress, we can approximate the evolution as being due to only one of the two effects at any given time. With this prescription, a cell behaves elastically and its elastic stress accumulates under the dominant Hall effect if $\tau < 1.1\tau_{el}$, the first term on the right side of Equation \eqref{dtau_dt}. When the cell reaches a critical value of $\tau = 1.1\tau_{el}$, it fails and behaves plastically. During its plastic phase, it releases the accumulated stress (the second term on the right side of Equation \eqref{dtau_dt}); the amount of release is fixed at 10\% of $\tau_{el}$. In the end, $\tau$ reduces down to 1$\tau_{el}$, corresponding to an energy release of $\sim10^{40}$ erg -- similar to an energetic short X-ray burst. 
The characteristic time scales ($t_{Hall}$ and $t_{pl}$) for $\tau$ evolution under Hall drift and plastic flow, then can be estimated as 800 and 9 yr respectively, using $\rho_e = 1.4 \times 10^{26}$ esu $\mathrm{cm}^{-3}$, $L = 200\,\mathrm{m}$, $\tau = 1.1 \tau_{el}$, and $\nu=10^{36}\,\mathrm{poise}$. However, typical durations for the Hall and plastic phases in the simulation are $\sim$ 75 years and 1 year, respectively. Since the reduction of $\tau$ is only 10\% of $\tau_{el}$ in the model, phase durations are shorter than the estimated characteristic time scales. Note that although the magnetic field strength does not appear explicitly in our formulation, our ansatz that the Hall phase builds stress in patterns that resemble our cellular lattice, and that Ohmic decay is subdominant and may be neglected, both mean that we are implicitly assuming a typical crustal magnetic field strength in the approximate range of $10^{14}-10^{15}\,\mathrm{G}$.

Previous studies \citep{Lander2019, Gourgouliatos2021} have shown that the elastic stress of a cell may exceed the yield stress by some tens of percent before any significant effect of the plastic failure. Note that this is also required in order to produce a sudden release of energy -- otherwise, a cell's evolution would saturate, with the plastic flow and Hall drift counteracting one another, to leave the stress at a constant value equal to the yield stress. Therefore, the model assumes an above-mentioned critical value of $\tau=1.1\tau_{el}$ for the failure of an individual cell. During its failure (plastic phase), its neighboring cells will be affected due to the exerted shearing force created by this plastic cell. To mimic this effect, the \cite{Lander23} model includes a cell rule that allows varying yield stress between 1 and 1.1$\tau_{el}$: a cell's yield stress is lowered by 0.025$\tau_{el}$ from 1.1$\tau_{el}$ for every plastic neighbor, assuming its behavior is only affected by its four neighbors. With this choice, if a cell has the maximum number of neighbors in the plastic phase, i.e., four, it enters the plastic phase as soon as its own stress reaches $1\tau_{el}$. Clearly, changing the precise values involved in the yielding process will result in quantitative differences in typical burst sizes. However, the main feature of the cell rule is robust: that nearby plastic flow expedites a cell’s failure, and the simultaneous plastic phase of adjacent cells enables the model to produce high-energy magnetar activities like flares.

A cluster of adjacent plastic cells is considered a single but larger cell with a plastic flow through the entire cluster at an average velocity,  $\overline{v}_{pl} \propto (\tau - \tau_{el}) / \nu$. This plastic flow within a plastic cluster causes a braiding of external magnetic field lines due to their footpoints embedded in the cells, increasing the average twist ($\psi$) of the associated coronal loop, $d\psi/dt = \overline{v}_{pl}$. It also represents the rate of energy transfer from the crust to the corona, $E_{clus} = \psi \times E_{max}$, where $E_{max}$ is the maximum expected energy considering all plastic cells in a cluster. The model simply assumes that when a plastic cluster ceases to exist, its associated energy is emitted as a single burst. Here, it is important to note that the transfer of elastic energy to burst energy released in high-energy photons is a highly complex process that may not be efficient (see e.g. \cite{belo13}). Given that we do not touch on this at all, our model is effectively degenerate in how much stress is relieved from the crust in a failure event, and the radiative efficiency of the emission process: an observed burst of energy $10^{40}$ erg could represent a 100\% efficient transfer of $10^{40}$ erg of elastic energy, or a 10\% efficient process following an initial crustal energy release of $10^{41}$ erg.

At the start of each simulation, the temperature ($T$) of all cells is set to a minimum `ambient' value of $5 \times 10^8$ K. During the plastic phase of a cell $T$ increases due to plastic flow, and then returns to the ambient temperature linearly over 13 yr starting from the end of the plastic phase; in both cases the treatment is as in \cite{Lander23}. At the end of the plastic phase, $T$ reaches a broad peak (for roughly a year) around its peak value, which is at most $2.5\times 10^9\,\mathrm{K}$. Note that temperatures much higher than this could not be maintained anyway, since the crust is cooled efficiently by neutrino emission from plasmon decay (see e.g. Figure 1 of \cite{yakovlev01}. The viscosity ($\nu$) of the crustal matter is inversely proportional to $T$, and its possible range during a simulation is between 10$^{34}$ and 10$^{36}$ poise. Since the newborn magnetar crust is unstressed and hence has no seismic activity, the time t$=$0 yr represents a highly-stressed young magnetar crust at age $\sim$1000 yr in this study. Therefore, stress levels of the crust cells are assigned randomly between 0.9 and 1.1$\tau_{el}$ at the start of each simulation, and hence they are all in the Hall phase at t$=$0. Finally, we note that there are $56\times14 = 784$ crustal cells covering the active Northern hemisphere of the magnetar, where crustal motion drives the increase in coronal twist; modeling half of the surface avoids conceptual issues related to whether an opposing crustal motion in the other hemisphere might annul coronal twist.


\subsection{What causes clustering of small bursts?} \label{sec:small_bursts}

The basic picture of neutron-star crustal energy release is that elastic stresses build up unobserved over the star's lifetime, until the crust's yield stress is reached, at which point stress is relieved through a failure of the crust, probably taking the form of a plastic flow. This flow will cause some localized heating, which could potentially be seen as an outburst: an additional thermal component of the X-ray spectrum that cools over some months. It does not, however, directly power the hard X-ray bursts that are our main focus here. Instead, the flow advects the footpoints of the magnetosphere, which twists up coronal loops. The presence of coronal loops may be seen indirectly by affecting the star's spindown, or directly when these loops undergo magnetic reconnection, which returns them to a lower-energy and less-twisted state; the excess energy is converted to an abrupt emission of high-energy photons together with, in some cases, a localized `trapped fireball' that decays more slowly, as seen most notably following giant flares \citep{TD01}.

A one-off burst does not itself provide stringent constraints on this sketch of how the magnetar's crustal energy becomes a photon count we can detect, but interpreting a cluster of bursts reduces the freedom in our modeling. As an example, let us consider the spectacular burst forest of SGR $1900+14$ on 29th March 2006, first analyzed by \cite{Israel08}. Their Figure 1 shows that over half a minute, the magnetar emits several protracted bright bursts with long tails, as well as a number of shorter and less energetic bursts. \cite{Israel08} refer to the former events, with higher energy and longer duration, as `intermediate flares', but conclude that there is no firm distinction between those and short bursts; we will simply refer to all as `short bursts'. This earlier study only reports the total energy of the burst storm, $\sim(2-3)\times10^{42}\,\mathrm{erg}$, but for a physical interpretation of the burst mechanism, it is helpful to have individual energies. We estimate that the brightest bursts have energies of $\sim(0.5-1.5)\times 10^{41}\,\mathrm{erg}$ in 10-100 keV by assuming a distance to the source of 10 kpc. The initial rise of a burst -- when it can be discerned -- is quasi-exponential over $\sim 25-50$ ms, and the longer bursts feature a pseudo-linear decaying tail of duration $\sim 1$ s. The brightest bursts typically come $\sim 2-4$ seconds apart.

We can now interpret the features of the burst forest of SGR $1900+14$. The onset of each burst is likely to be the result of magnetic reconnection of a twisted coronal loop, with the protracted nature of the brighter bursts representing a small trapped fireball attached to the star. Each burst represents a local failure, since a large-scale event would both be more energetic and protracted enough in time to show rotational modulation (as seen in the decaying X-ray tails following the giant flares of SGRs $1900+14$ and $1806-20$). Furthermore, the energy release from the larger bursts $\sim 10^{41}\,\mathrm{erg}$ is comparable to the maximum expected from the failure of a single $1\,\mathrm{km}^2$ crustal cell, so it is not possible to explain the whole $\sim 30$-second period of activity as a single twisted coronal loop undergoing repeated reconnection events, nor would persistent plastic flow in a single cell be fast enough to `recharge' the coronal loop with energy for a new burst, since the timescale for this process is likely around a month; see \citet{Younes2022}.

This leaves us with a scenario where several highly stressed local regions of the crust contribute to the burst storm. The first of these localized failures exerts a torque that increases the stress in the surrounding region, and will also send shear waves with speed $\sqrt{\mu/\rho}$ (where $\mu,\rho$ are the shear modulus and mass density, respectively) across the star's crust. The latter mechanism could, in principle, trigger distant crustal failures at any highly-stressed point in the crust, but these would lead to subsequent events being separated by at most the shear-wave crossing timescale for the crust, $\sim 2\pi R_*/\sqrt{\mu/\rho}\approx 1$ second (where $R_*$ is the stellar radius, which we take as 12 km). Instead, we see somewhat longer and non-regular waiting times between bursts, suggesting that the interaction between plastically-failing cells and their surroundings may be driving the clustering. We wish to investigate whether our earlier cellular-automaton model, which aims to encapsulate magnetar activity with minimal assumptions, is able to explain this burst clustering.


\section{MODEL \& SIMULATION} \label{sec:model}

The model outlined in \cite{Lander23} allows for magnetar activity across a full range of energies, corresponding to those of short bursts and intermediate \& giant flares, all of which are powered by the release of stress from a crust that behaves as a cellular automaton and drives coronal and then bursting activity. The motivation for the original code was to model the energy output from a magnetar on long timescales rather than to resolve short-timescale phenomena specific to any particular magnetar, but the model can readily be extended and adapted to study many magnetar activities. Our study aims to understand the nature of short magnetar bursts by focusing the simulation on high-energy short bursts of magnetars. To do that, we aimed to match the model results with short magnetar burst observations. While doing so, we ensured that all new implementations to the code were physically motivated and meaningful. In the following, we describe the four differences between the present implementation and that of \cite{Lander23}, each detailed in one paragraph.


To study short bursts, we focus on shallow failures of the cells with a depth of 0.2 km (a density range $4.0\times 10^{11}\,\mathrm{g}\,\mathrm{cm}^{-3}<\rho<1.5\times 10^{13}\,\mathrm{g}\,\mathrm{cm}^{-3}$). This means that a cell does not fail deeply, down to the crust-core boundary. The original version of the simulation also allowed these deeper failures to explain the larger energy activities, such as intermediate and giant flares. 
This was intended as a way to model whatever additional physics might -- in rare instances -- cause a larger spreading failure across the crust, in whose wake the magnetar would become significantly less active (as seen in the aftermath of the three giant flares observed locally, to date). Clearly, however, the rarity of giant flares indicates that their trigger may only occur once in several decades or more, whereas here we are interested in the far more common small bursts, and their clustering over short timescales. It does not make sense, therefore, to implement the previous deep failure criterion in this case. 


We also changed our assumption about how energy is released from a group of neighboring cells. In the previous letter, it was assumed that contiguous regions of plastic cells all contributed their energy to a single large-scale coronal loop. This approach was also needed to explain higher-energy flares, since even the deep failure of a single cell does not reach the $\gtrsim 10^{44}\,\mathrm{erg}$ energy range needed to explain a giant flare. Here, we therefore assume that each cell releases energy to the corona individually, which allows the possibility of several successive, distinct bursts occurring in a local region due to cell interactions.


As mentioned in Section \ref{sec:sam_model}, 
crustal magnetic field lines that are dragged around in a cell due to its circulating plastic flow exert a shearing force on its neighboring cells, therefore, nearby plastic failure quickens a cell failure. While encoding this effect, the motivation for the cell rule in \cite{Lander23} is to allow simultaneous plastic phases of many adjacent cells in order to model the rarely-observed giant flares. Here, we focus on the far more frequently observed short bursts, and from magnetar observations, we know that these do not usually occur in isolation but rather successively. To mimic this behavior of magnetar bursts, we need to make a cell more sensitive to its neighbors' behavior. To achieve this in our simulation, we modified the cell rule: If a cell has one or more (2, 3, or 4) plastic neighbors, it fails for any stress above $\tau_{el}$, otherwise, it fails at $\tau = 1.1\tau_{el}$. With the modified cell rule, we increase both the number of cell interactions and the number of successive failures of a cell compared to the original version. In general, when a cell fails and enters a plastic phase, it has at least one plastic neighbor and fails successively usually twice, and at most $\sim$5-6 times. 

\begin{figure}[htbp!]
    \centering
    \includegraphics[width=0.7 \textwidth, trim=65 290 75 265, clip]{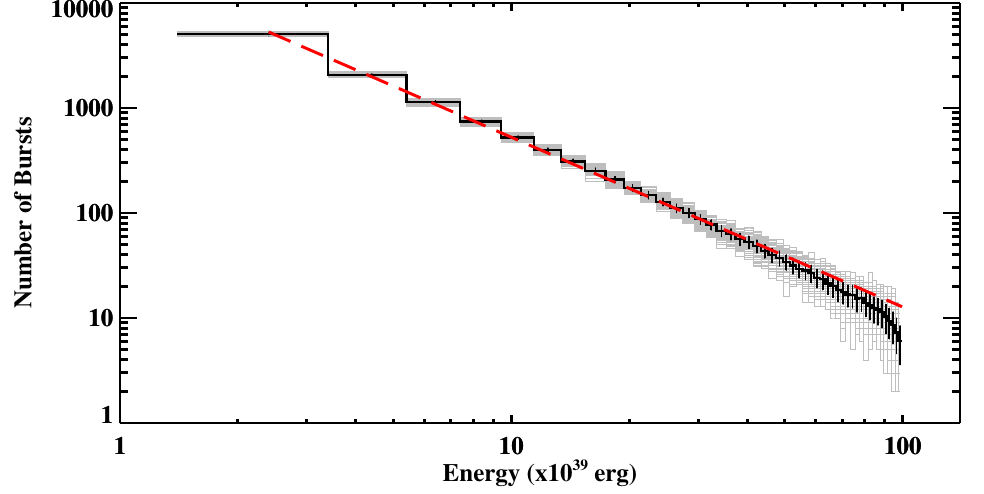}
    \caption{The distribution of burst energies obtained by averaging the results of 100 simulations (solid black lines). The red dashed line shows a PL fit to the average distribution. The gray lines display the distributions of individual simulations.}
    \label{fig:burst_en_dist}
\end{figure}

When a cell fails, we do not expect it to release all its stress, but rather a portion of it. In the original model, this portion was a fixed value, namely a 10\% reduction. However, an arbitrary amount of stress loss is more likely and physically expected. Note that the amount of stress reduction is directly related to the amount of energy transferred to the corona via braiding of the magnetic field lines due to plastic flow during failure. From the observational point of view, less energetic bursts are more frequently observed, and so more abundant than the higher-energy ones. Therefore, when a cell fails, instead of a fixed stress loss ($\tau_{loss}$), we assumed a random $\tau_{loss}$ between 0.01$\tau_{el}$ and 0.6$\tau_{el}$ drawn from a power law (PL) distribution with an index of 1.6. This allows us to cover an energy range of $10^{39}-10^{41}$ erg, which is the typical energy range of recursive short bursts. More importantly, this approach enables us to obtain a burst energy distribution that follows a PL trend whose index is 1.6, consistent with the burst energy distribution of real magnetar burst observations \citep{EG99, EG20}. Figure \ref{fig:burst_en_dist} shows the burst energy distribution of the model star through the 1000-year evolution of all crust cells. We found that the distribution follows a PL trend whose index is $1.60\pm0.01$ with its 1$\sigma$ uncertainty. 



\section{Results}\label{sec:res}

Including the above-mentioned implementations, we simulate short magnetar burst activities for a fiducial simulation duration of 1000 yr. As a reminder, t=0 yr represents a highly stressed young magnetar whose age is $\sim$1000 yr in our simulation. This is enough time to collect reliable statistics for burst clustering that vary little from simulation to simulation. Having experimented with longer simulations, we found these give no additional information, as the cellular automaton behavior saturates in well under 1000 yr, and because there is no additional secular evolution of the system on long timescales. Besides, observationally, burst prolific magnetars have low characteristic ages, at most a few kyr, and our motivation is to investigate these highly stressed young magnetars in their burst active states.

\begin{figure}[ht!]
    \centering
    \includegraphics[width=0.7 \textwidth, trim = 75 160 75 340, clip]{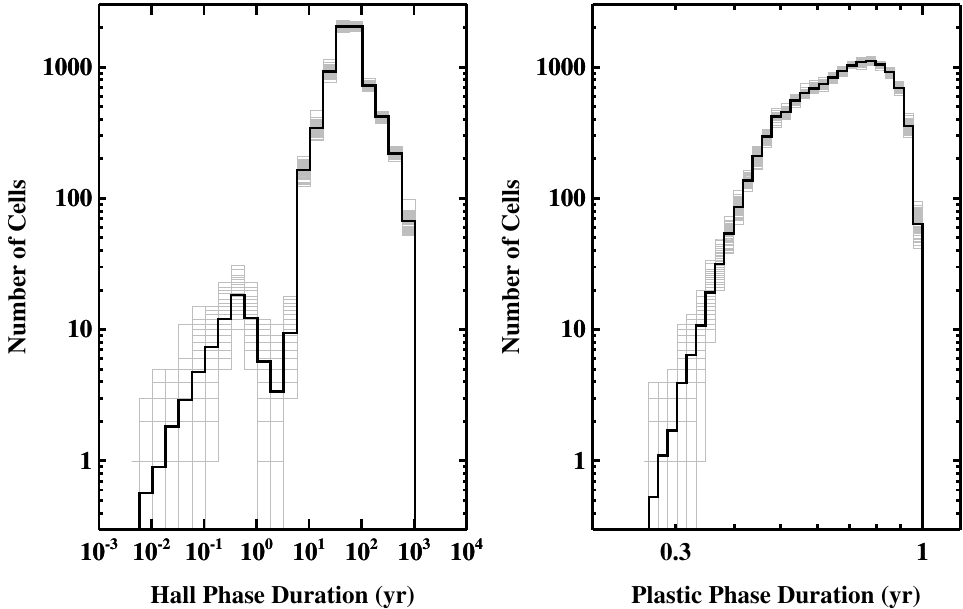}
    \caption{The distribution of Hall (left panel) and plastic (right panel) phase durations. Both trends shown with solid black lines were obtained by averaging the results of 100 simulations. The gray histograms display the distributions of individual simulations.}
    \label{fig:phase_dur_dists}
\end{figure}

Figure \ref{fig:phase_dur_dists} presents the duration distribution of the Hall (left panel) and plastic (right panel) phases of cells. Both Hall and plastic phase durations vary in our simulations due to a random amount of stress release rather than a fixed one (e.g., 10\%) during a cell's failure. A Hall phase generally lasts between $\sim$1 and 1000 years, with an average of 50 years. It occasionally lasts less than a year;  these cases are due to cells that have lost a small amount of stress, down to just below their yield values. Therefore, their stress levels re-reach their yielding values more quickly. The relation between the amount of $\tau$ increase during a Hall phase and its duration is simple: the higher the stress increase, the longer the phase duration. On the other hand, there is a rather complex relation between the plastic phase duration and $\tau_{loss}$ during a plastic phase. The plastic phase duration depends on the $\tau$ values at which the plastic phase begins and ends, and viscosity (hence, the temperature of the failing cell). Yet, we have a well-constrained plastic phase duration between one-third of a year and one year (0.3 - 1 yr), with an average of 0.7 yr. Therefore, a second burst from the same cell can not be observed on shorter timescales than the plastic phase durations, even in the case of successive cell failures. 

\begin{figure}[ht!]
    \centering
    \includegraphics[width=0.45 \textwidth, trim = 80 80 85 265, clip]{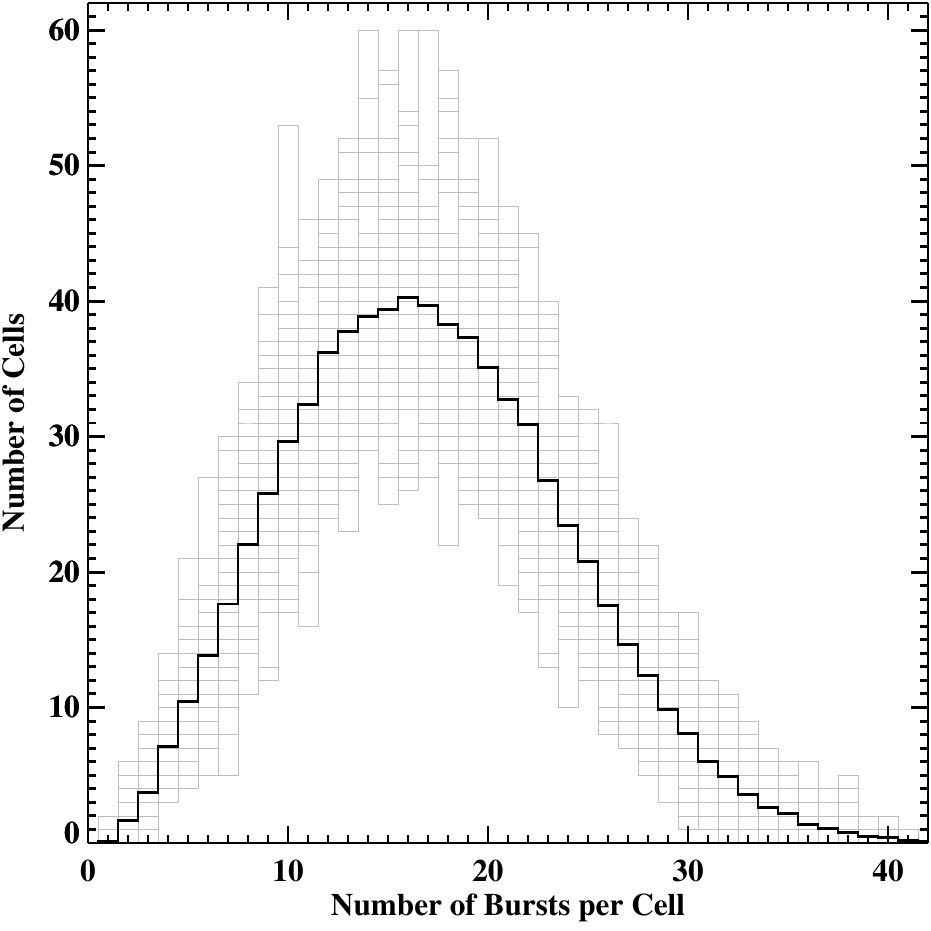}
    \caption{The distribution of the number of bursts per cell in 1000 years, obtained as an average of 100 simulations, is shown in black. The gray histograms show the distributions of individual simulations.}
    \label{fig:ind_cell_burst_dist}
\end{figure}
\begin{figure}[ht!]
    \centering
    \includegraphics[width=0.7 \textwidth, trim = 80 165 80 340, clip]{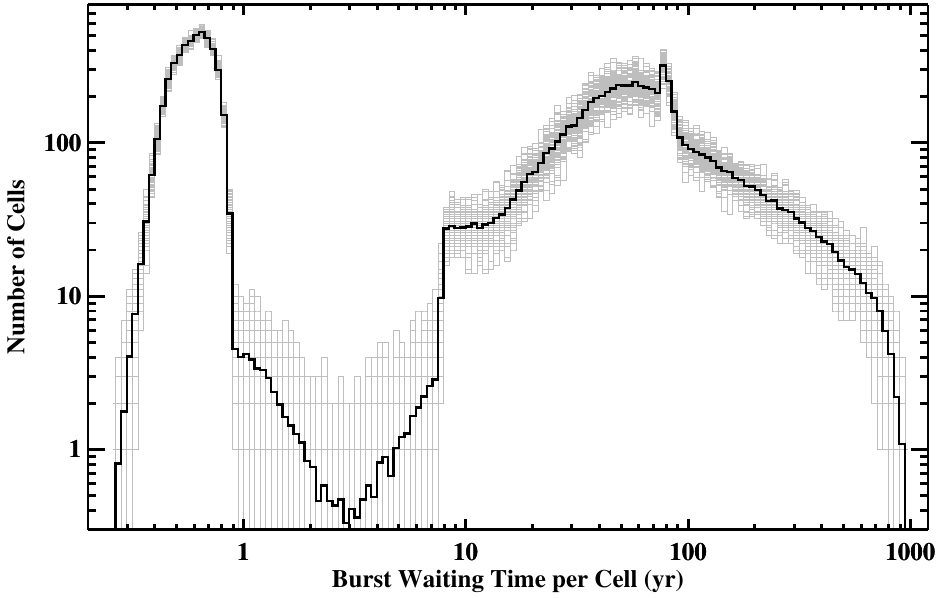}
    \caption{The distribution of waiting times between successive bursts from the same cell, considering all crustal cells. The solid dark line is the average distribution of 100 simulations, and the gray lines are the distributions of individual simulations.}
    \label{fig:tdiff_btw_bursts}
\end{figure}

Figure \ref{fig:ind_cell_burst_dist} shows the distribution of the number of bursts per cell. We find that a cell yields, on average, around 16 events. This is understandable given that the Hall phase lasts about 50 years on average, therefore, each cell releases energy to the corona $\sim$20 times over the course of 1000 years. We also calculated waiting times between successive bursts from the same cell, considering all cells throughout the entire crust. We present this distribution of waiting times in Figure \ref{fig:tdiff_btw_bursts}. 
The cells fail repetitively due to the cell rule (successive plastic phases of the cells), and since the plastic phase takes $\sim$ 0.3 - 1 yr, these cases create the data points around the first peak we observe in the figure. Non-successive cell failures create the large region around the second peak in the figure since these cases include a Hall phase in addition to a plastic phase between the two bursts, therefore, it requires a longer time for a cell to re-burst since a Hall phase is generally much longer than a plastic phase (see Figure \ref{fig:phase_dur_dists}). The small additional overabundance of bursts jutting out from this broad peak at $\sim 70-80$ yr corresponds to the numerous cases where a single cell with no plastic neighbors fails on its own at $1.1\tau_{el}$ and loses an average amount of stress from the power-law distribution and so drops back to $\sim 1.0\tau_{el}$, similar to the generic single-cell failure from the previous paper. In the most extreme cases, large stress reductions in the plastic phase mean that the subsequent Hall phase takes hundreds of years to bring that particular cell back to the point of failure and, if this is followed by another large plastic failure, it may only produce (say) two bursts over the 1000-year run of the code. 

\begin{figure}[ht!]
    \centering    \includegraphics[width=0.7\textwidth, trim = 35 117 25 250, clip]{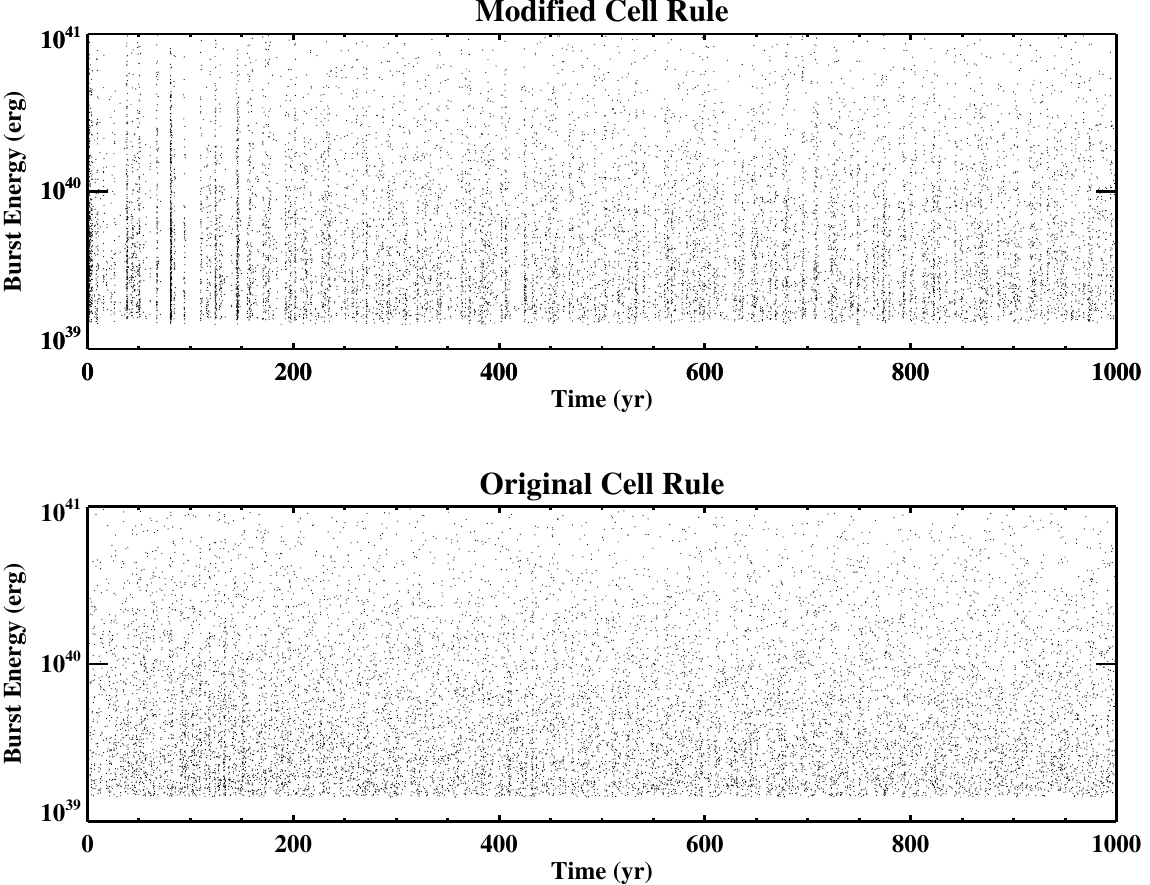}
    \caption{Energy of bursts over 1000 yr obtained with the modified cell rule (top panel) and with the original cell rule (bottom panel). Clustering is clearly more pronounced with the modified rule.} 
    \label{fig:burst_vs_time}
\end{figure}

\begin{figure}[htbp!]

\centering
    \includegraphics[width=0.58 \linewidth, trim=80 290 75 260, clip]{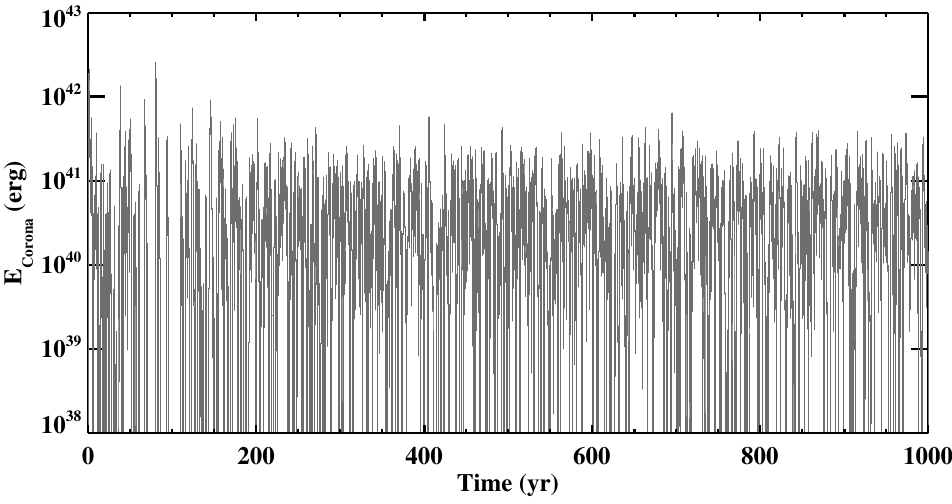}
\hfill
    \includegraphics[width=0.40 \linewidth, clip]{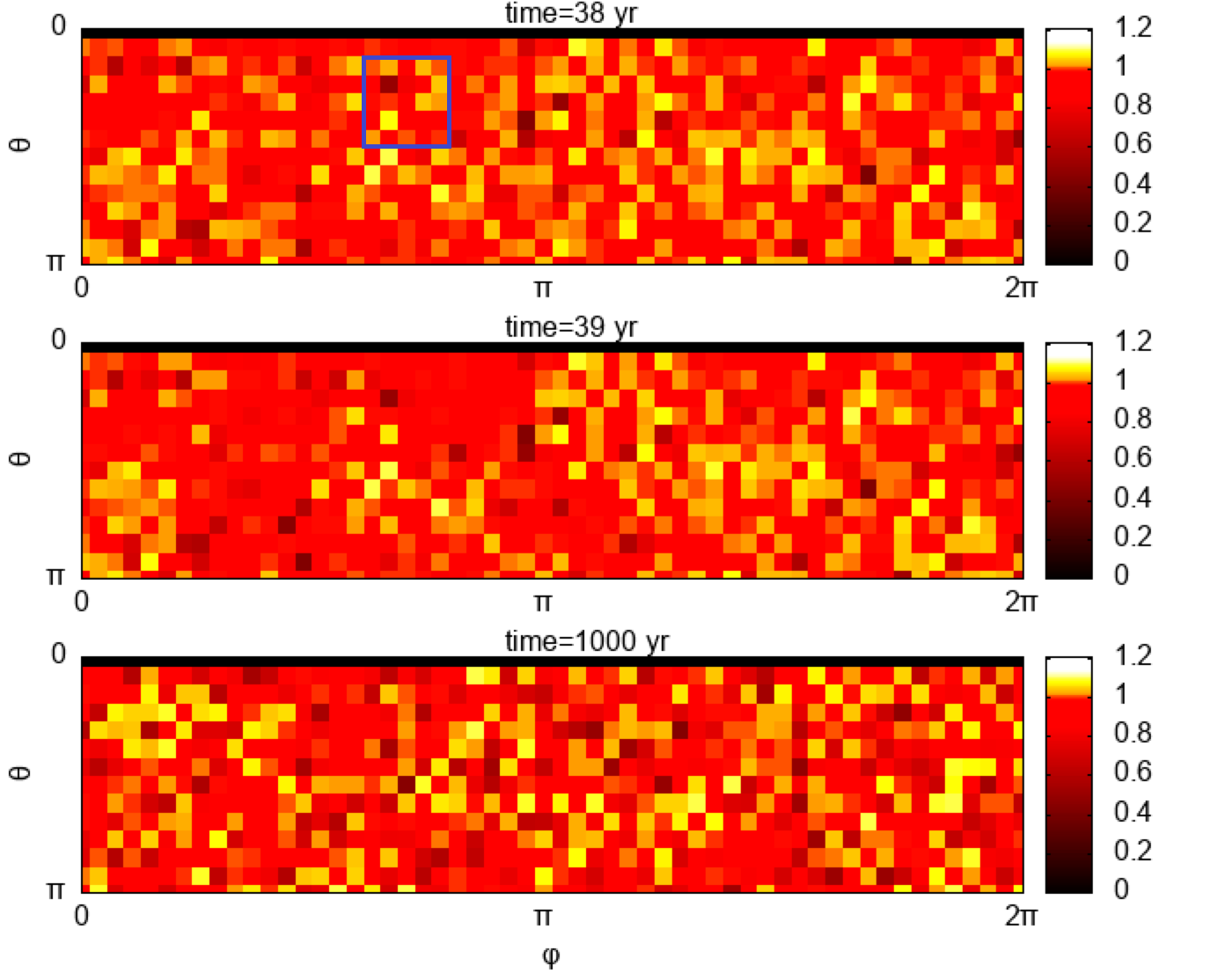}

\caption{[LEFT] The total twist energy of all coronal loops vs. time during the 1000-yr evolution with our modified cellular automaton. [RIGHT] Snapshots of stress evolution in units of $\tau/\tau_{el}$ (color scale) across the northern hemisphere, just before (top) and after (middle) the emission of a burst cluster with energy 2$\times10^{42}$ erg, and at the end of 1000 yr of evolution (bottom), for the same simulation as in the left-hand panel of this figure and the top panel of Figure \ref{fig:burst_vs_time}. The blue window shown in the top panel represents the highly stressed crustal region that triggered the successive failures.}
\label{fig:cell_evolution}
\end{figure}

As an important outcome of our new approach, we obtain clustered bursting episodes in time and present them in the top panel of Figure \ref{fig:burst_vs_time}. The bottom panel of Figure \ref{fig:burst_vs_time} shows the bursts obtained with the original cell rule to compare the two distinct bursting behaviors.
To enable direct comparison between the evolution shown in the top panel of Figure \ref{fig:burst_vs_time} and Figures 2 \& 3 of \cite{Lander23}, we next look at details of the coronal energy and crustal stress evolution for our modified cellular automaton simulation. In the left-hand panel of Figure \ref{fig:cell_evolution}, we plot the total coronal energy as a function of time over 1000 yr for our modified cellular automaton model; the main differences compared with the original approach are the lack of very bright ($\gtrsim 10^{43}\,\mathrm{erg}$) events, and the more frequent drops to zero coronal energy, leading to a more jagged-shaped line. In the right-hand panel of Figure \ref{fig:cell_evolution}, we show three snapshots of $\tau/\tau_{el}$: directly before and after the large burst cluster at time $\sim 38\,\mathrm{yr}$, and the final state after 1000 yr. As one of the cells in a highly stressed crustal region (blue window in the top panel) fails, it quickens the failure of its neighbors, shown in the window, and they all fail together and start to affect other neighboring cells. The burst cluster can be seen to result in a large region of relatively low stress (large red patches in the middle panel), but -- unlike the previous paper's deep failure criterion -- this is not permanent, and no trace of it can be seen in the 1000-yr snapshot (bottom panel). 

\begin{figure}[ht!]
    \centering
    \includegraphics[width=0.5\textwidth, trim = 80 95 55 260, clip]{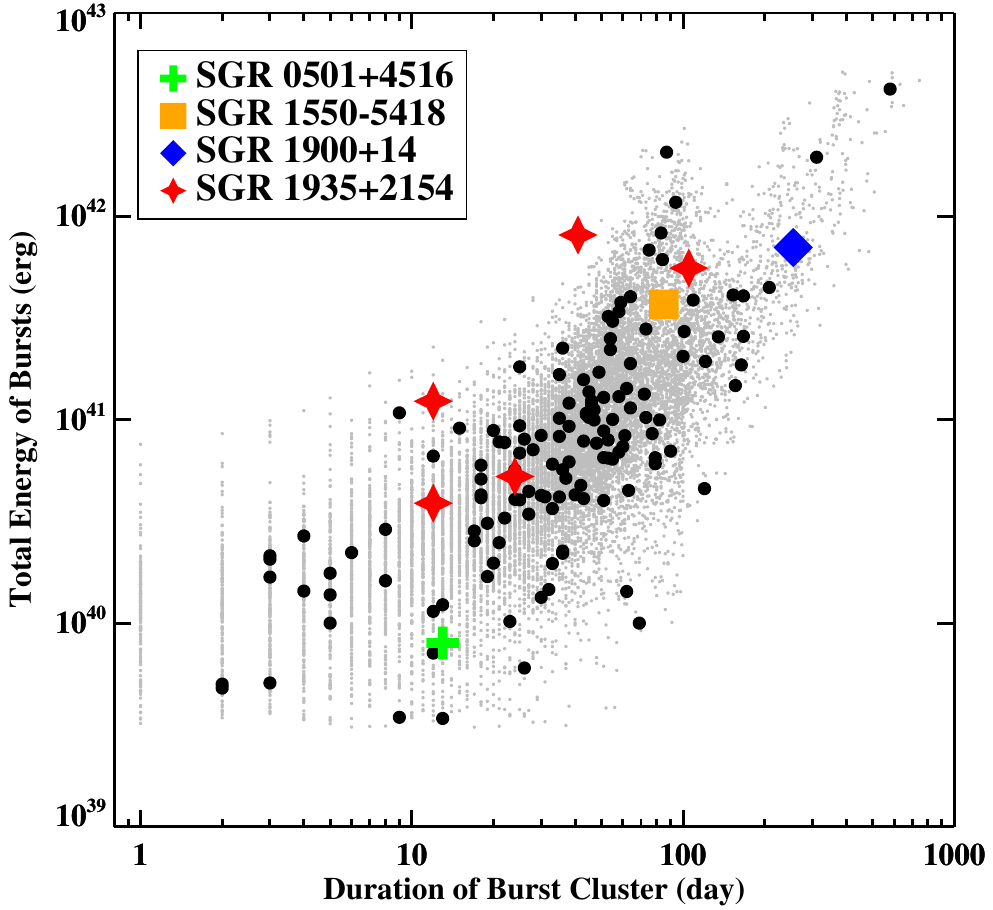}
    \caption{The plot of the total burst energy vs. duration of each cluster from 100 simulations (grey dots). The black circles show the results of a single simulation, the one shown in the top panel of Figure \ref{fig:burst_vs_time}. Colored data points represent the same relation for the actual observations of magnetars: SGR $0501+4516$ (green cross), SGR $1550-5418$ (orange square), SGR $1900+14$ (blue diamond), and SGR $1935+2154$ (red stars).}
    \label{fig:burst_clus}
\end{figure}

The clustering of bursts in time is a direct indicator of burst-active episodes of magnetars. However, defining the interval of burst clusters within a simulation is not straightforward. In our set of simulation runs, the number of bursts varied between around 12,000 and 13,000, with an average of $\sim$12,500 bursts, of which $\sim$10,000 are single bursts in the days they occurred. The rest of the bursts ($\sim$2000) accumulated such that there are 2-10 bursts in a single day. In other words, there was no activity during $\sim$355,000 (97\%) days in 1000 yr. To define a burst cluster, we grouped the burst-populated days based on having 2 or more bursts. Time intervals with at least 100 subsequent days of no activity (2 or more bursts) are considered burst-quiescent episodes. Therefore, the duration of a burst cluster is then determined as the time from the activity onset to the beginning of the following burst quiescent interval. In Figure \ref{fig:burst_clus}, we present the duration of burst clusters vs. the total energy of bursts within corresponding clusters. We note that the durations of burst clusters lie on the integer values in the figure since we used a time resolution of one day while binning the data.

To compare the duration and energetics of burst clusters with actual magnetar bursting activities, we utilized time-integrated spectral studies of recursive short magnetar bursts in the literature. Using the above-mentioned burst cluster definition, we could identify burst clusters for four magnetars (see Table \ref{tab:MagnetarList}). For SGR 0501+4516, SGR 1550-5418, and SGR 1935+2154, we used the burst energy fluence values in 8-200 keV observed with \fermi and calculated the isotropic burst energies assuming distances to the sources provided in the table. Similarly, we obtained the total energy released from the short bursts of SGR 1900+14 by using the burst fluence values in $>$ 25 keV observed with \cgro and the distance to the source given in the table. Overplotting the durations and energetics of burst clusters from the observed magnetar activities in Figure \ref{fig:burst_clus} reveals a convincing agreement with our simulation results. To quantify this consistency, we obtained log-normal distributions of the total burst energies from 100 simulations (gray dots), corresponding to the burst cluster durations of actual magnetar bursting activities. We found that most observational data shown in Figure \ref{fig:burst_clus} remained within about 2$\sigma$ level of the associated energetics distributions. We note that only one burst cluster of SGR 1935+2154 (which occurred in April-May 2020) corresponds to the observed energetics that is slightly exceeding the 3$\sigma$ level.

\begin{deluxetable*}{lcccc}
\tablecaption{A Sample of Magnetars with Extensive Burst Active Phases}
\tablewidth{0pt}
\tablehead{
\colhead{\bf Source Name} & \colhead{\bf Activity Period}  & \colhead{\bf Distance} & \colhead{\bf Total Energy of} & \colhead{\bf Reference} \\
 & & \colhead{\bf (kpc)}  & \colhead{\bf Bursts (10$^{39}$ erg)} & } 
\startdata
\hline
SGR 0501+4516 & August-September 2008 & 2 & 8 & 1, 2\\
\hline
SGR J1550$-$5418 & January-April 2009 &  5 & 370 & 1, 3\\
\hline
SGR J1900+14 & May 1998 - February 1999 & 10 & 703 & 4, 5\\
\hline
& February-March 2015 & & 39 & 6\\
& May-August 2016 & & 557 & 6\\
SGR J1935+2154 & November 2019 & 9 & 123 & 7\\
& April-May 2020 & & 810 & 7\\
& January-February 2021 & & 53 & 8\\
\enddata
\tablecomments{\\
References: (1) \cite{Collazzi2015}; (2) \cite{Xu2006}; (3) \cite{Tiengo2010}; (4) \cite{EG99}; (5) \cite{Israel08}; (6) \cite{Lin2016}; (7) \cite{Lin2020}; (8) L. Lin et al. 2025, in preparation
}
\end{deluxetable*}\label{tab:MagnetarList}

Finally, we note that we performed more simulations with different initial stress levels and different cell rules, to check whether our results were generic outcomes of the simulation, or instead had some strong dependence on the specific choices we made. We still obtained a similar clustered bursting behavior of short bursts, confirming the robustness of our results (see Appendix \ref{appA}).


\section{Discussion}\label{sec:dis}

Burst clusters are among the most spectacular activities of magnetars, and -- unlike the ultra-rare giant flares -- are relatively common. Despite this, to our knowledge, there has been little attempt to provide a theoretical interpretation of this phenomenon. \cite{li_etal16} studied Hall waves that are emitted during each crustal failure with a 1D model. These waves propagate and trigger the failures of other regions as a potential source for magnetar outbursts, which are long-lasting (weeks to many months) persistent X-ray flux enhancements \citep{CotiZelati}. The bursts in the focus of this study, on the other hand, are of short-duration hard X-ray events. Whilst our model does also include a plastic heating term coinciding with the short bursts, which broadly resembles the physics of an outburst, we have not attempted any detailed modeling of this process.

Having experimented with some minor modifications to a previous cellular automaton model for the magnetar crust \citep{Lander23}, we have found that we can reproduce the basic features of real magnetar burst clusters. We stress, however, that our success in matching observations is not a result of fine-tuning of many variables; there are only a limited number of adjustable parameters or prescriptions in this model, and that the results we obtain are robust to quantitative changes in these parameters (see Appendix \ref{appA} for some tests of this). With the agreement between our model results and real magnetar observations, we can now invert the reasoning to infer details of the physics of the magnetar crust and to consider when this paper's model is applicable.

One change implemented here was to allow the reduction in stress in a cell's plastic phase to be taken randomly from a power law distribution rather than a value fixed by the number of plastic neighbors the cell has. Because cell stress relief directly corresponds to burst energy in our model, this allowed us to produce short bursts over a realistic energy range corresponding to magnetar observations. However, it also provides a very plausible physical picture of the crustal failure process: local anisotropies due to, e.g., seismic history and failure dynamics are indeed likely to lead to variable stress relief in a plastic phase (see also \cite{li_etal16}).

Within our model, a burst cluster can arise as a result of successive short plastic phases of the same cell or different cells triggering one another. Our model simply assumes that all energy transferred to the corona is emitted at once at the end of a plastic phase, since it does not include emission physics, and so both kinds of `burst' look similar in our simulations. More realistically, this energy would be temporarily stored in twisted coronal loops and then released due to magnetic reconnection in one or more events. Following the reconnection, synchrotron-like nonthermal radiation spectra are expected from these events due to particle interactions in the magnetosphere. Here, it is unknown whether the energy transferred to the corona from a single cell will create a single burst or multiple bursts. However, from the arguments at the start, we do not expect repeated large bursts, since each cell can release total energy no greater than $\sim 10^{41}\,\mathrm{erg}$.

As can be seen in Figure \ref{fig:burst_clus}, the duration of a burst cluster from our simulations is mostly in the range of 10-100 days. We know that the minimum amount of time required to obtain a second burst from the same cell is given by the minimum duration of the plastic phase, which is one-third of a year ($\sim$ 120 days; Figures \ref{fig:phase_dur_dists} and \ref{fig:tdiff_btw_bursts}). Therefore, bursts in a cluster are mostly due to the failures of different active cells. This can be seen visibly in maps of the crust's stress: directly after any large burst cluster, a substantial contiguous low-stress region can be discerned (see the right-hand middle panel of Figure \ref{fig:cell_evolution}). We can interpret this as any highly stressed region being able to affect its neighbors quickly, and hence make the magnetar burst-active. Although a single active coronal loop is theoretically possible, we generally do not observe any coronal loop that twists repeatedly within the duration of a burst cluster in our simulations, considering the required time interval for a plastic flow to re-create enough twists for a second reconnection, unless a burst-active episode lasts long enough to create multiple bursts from the same cell (assuming successive and small failures). Therefore, most of the successive events we identify here as `burst clusters' are due to active neighboring cells from an active region(s). Then, successive small failures of a single cell may lead to consecutive clusters of bursts with short quiescent periods. This can be interpreted as the behavior of magnetars that become almost annually active, such as SGR J1935+2154. Since the remaining stress of the cells is still high enough, they can reach their yielding stress quickly and become active repeatedly until they release enough elastic stress to be silent for a longer period. 

There is a compelling similarity between our model and observations of burst energy output. In Figure \ref{fig:burst_clus}, we plotted total burst cluster energy output as a function of duration, finding that both simulated and real data exhibited the same strong correlation. This suggests that although burst clusters represent episodes of significant seismic activity, the luminosity during a cluster is relatively constant (however long it lasts), showing that short-burst events often trigger other short-burst events, but do not cause a runaway effect leading to more violent phenomena. In particular, given that the total elastic energy reservoir in a fully-stressed magnetar crust is over $\sim 10^{46}\,\mathrm{erg}$ \citep{lander15, Lander23}, even the most protracted burst clusters are unlikely to release more than one thousandth of the total crustal energy reservoir. This strongly suggests that producing a giant flare requires one distinct additional physical ingredient, beyond those in the model discussed here. The previous paper \citep{Lander23} tried to encapsulate this additional physics in a `deep failure' cell rule, but without pinpointing what exactly that would mean physically.

The previous paper regarded collections of plastic cells neighboring one another as a single entity: a region where the plastic flow circulated around the entire domain, twisting up a geographically large-scale coronal loop whose energy was considerably larger than that associated with the coronal loop of a single cell. This characterization was useful for producing highly energetic `events' in the approximate energy range of $10^{43}-10^{45}\,\mathrm{erg}$.
However, it also prevents clusters of bursts from occurring due to neighboring cells interacting, since those neighbors would simply join to make a single larger burst -- and so we ceased to use that rule here. Instead, the physical picture we have is of a more complex corona, with potentially large numbers of physically distinct, thin coronal loops, of diameter similar to that of the cell they emerged from, $\sim 1\,\mathrm{km}$.

With the rarity of giant flares and our present focus on more typical events on shorter timescales, there was no reason to enable the additional, complicating feature of deep failures from the previous paper (which were assumed to occur when a critical number of neighboring cells were plastic). Instead, however, we have a greater sensitivity to failure if a cell has just one plastic neighbor. This suggests that even if a magnetar has suffered a giant flare that reduces stress across large patches of its crust, it would still be able to power burst storms even from the small, isolated active regions that remain; this provides a physical picture for the 2006 burst storm of SGR $1900+14$, $7\scriptstyle{\frac{1}{2}}$ years after its giant flare \citep{Israel08}.

This paper has tried to take a step towards a firmer model of the mechanism driving magnetar activity, focusing on how different regions of a highly-stressed crust may communicate with one another to produce clusters of failure events. It is clear that the model remains qualitative in nature, with some details of the physics necessarily glossed over in the use of cells and interactions between them. Regardless of the specifics of our model, however, we suggest that any viable mechanism for producing both single bursts and burst clusters must allow for local regions of the crust to be semi-autonomous most of the time -- resulting in localized failures and single bursts -- but to interact with other parts of the crust under certain conditions, thus driving the more energetic burst-cluster events. Here, we model clustering as being related to direct communication between nearby parts of the crust. Other -- potentially more complicated -- possibilities might be interaction via the corona \citep{Younes2022} or the core \citep{Thompson2017}. Some logical next steps for our work would be to try to put the cell interactions on a sounder theoretical footing through first-principles simulations of crustal failure in a strong magnetic field, as well as detailed consideration of the coronal dynamics, including reconnection. Nonetheless, the broad agreement between the model and observations so far suggests that this is a good foundation on which to build.


\section*{acknowledgments}
We thank the anonymous reviewer for reading the manuscript carefully and providing precise and constructive comments. This work was supported by a Royal Society International Exchange grant IES$\backslash$R3$\backslash$223220 between the University of East Anglia and Sabanc\i\ University.


\appendix
\section{Robustness tests of the Simulation Results} \label{appA}

\begin{figure}[ht!]
    \centering
    \includegraphics[width=0.7\textwidth, trim = 35 90 25 235, clip]{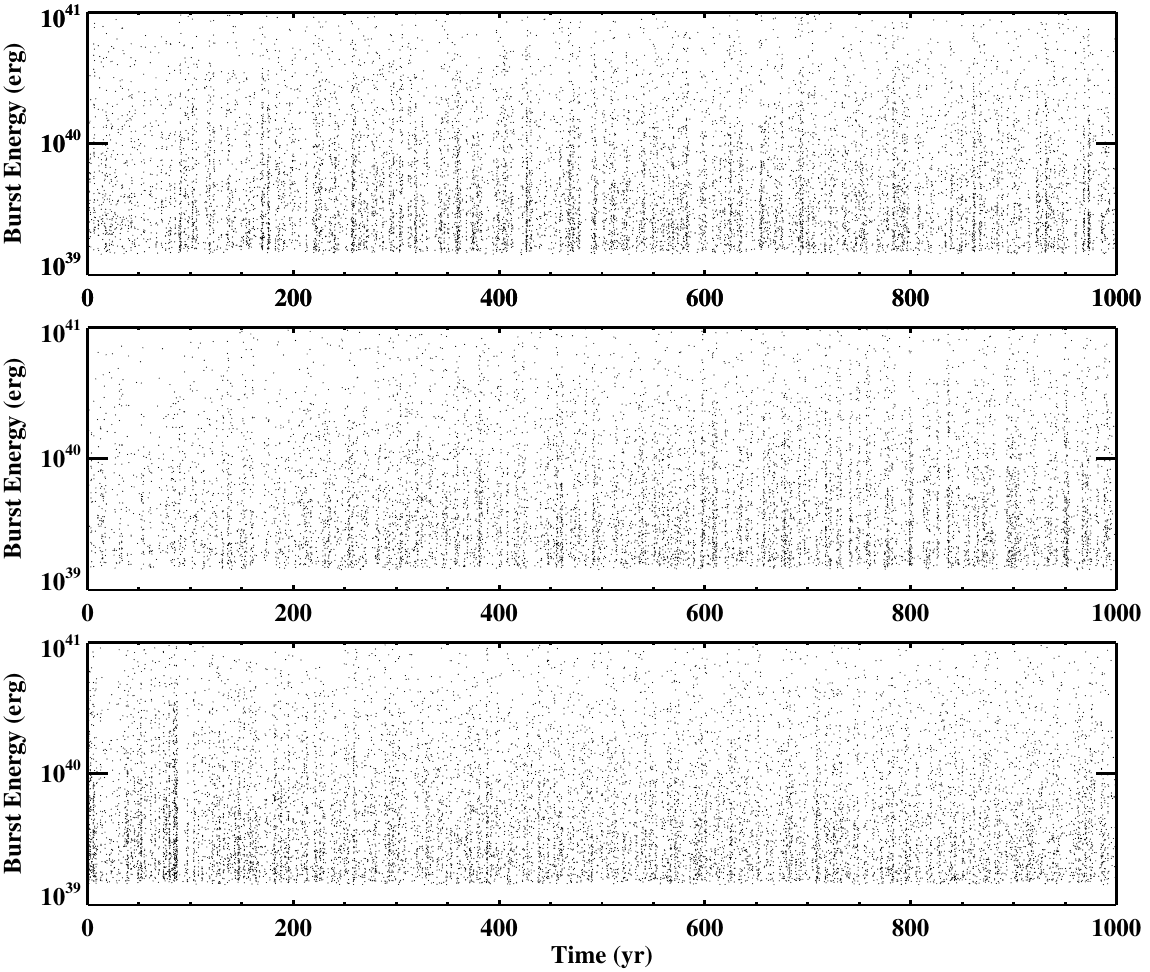}
    \caption{The burst energies over a 1000 yr obtained with [TOP] Random initial stress values between 0.7$\tau_{el}$ - 1.1$\tau_{el}$. [MIDDLE] Random initial stress values between 0.5$\tau_{el}$ - 1.1$\tau_{el}$. [BOTTOM] Another cell rule (see text for details).}
    \label{fig:burst_vs_time_2}
\end{figure}

\begin{figure}[htbp]
    \centering
    \includegraphics[width=0.32 \textwidth, trim = 55 75 50 255, clip]{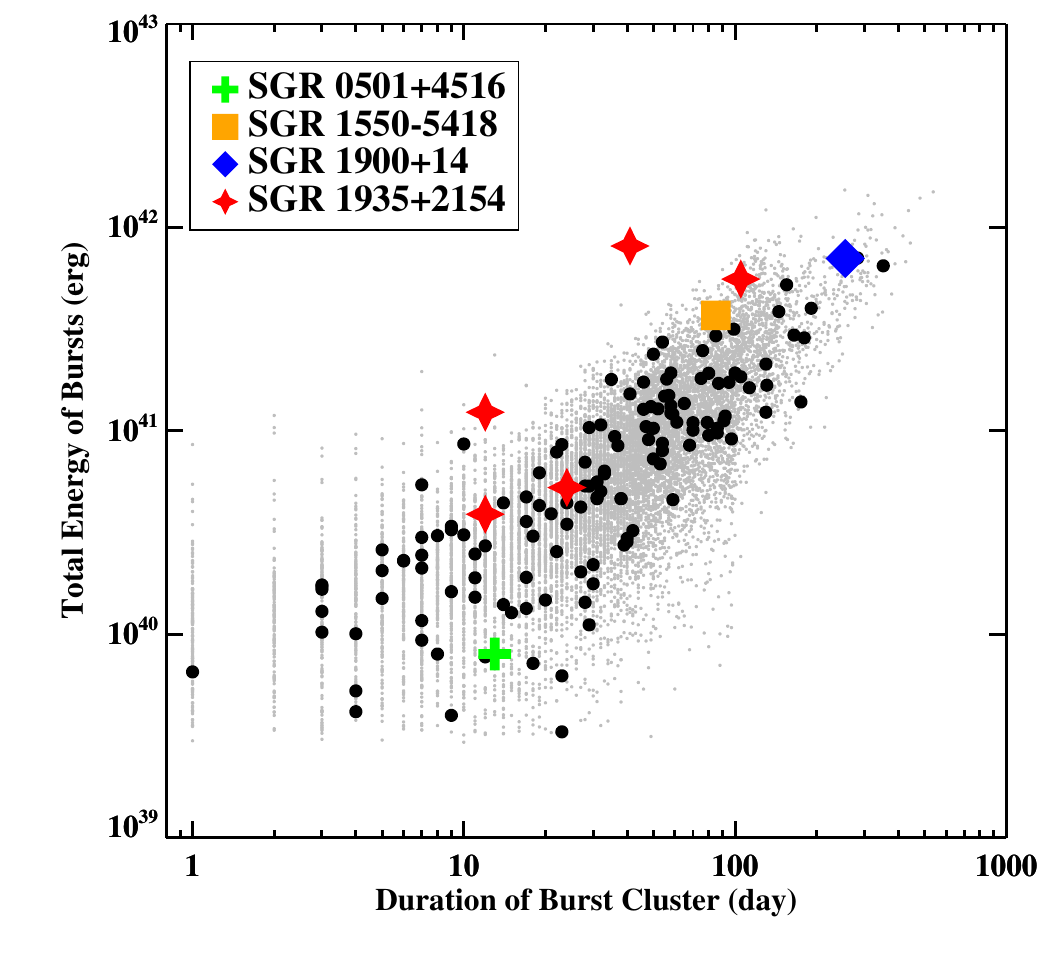}
    \hfill
    \includegraphics[width=0.32 \textwidth,  trim = 55 75 50 255, clip]{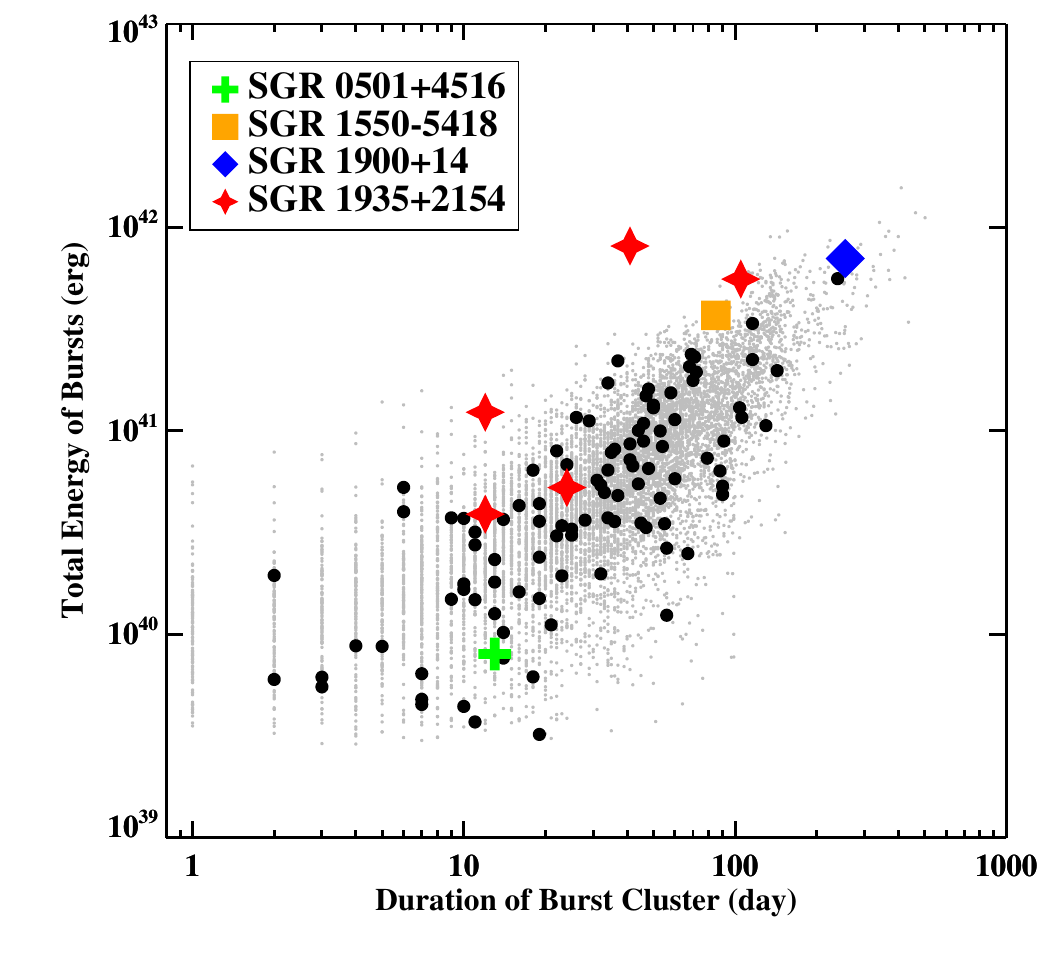}
    \hfill
    \includegraphics[width=0.32 \textwidth, trim = 55 75 50 255, clip]{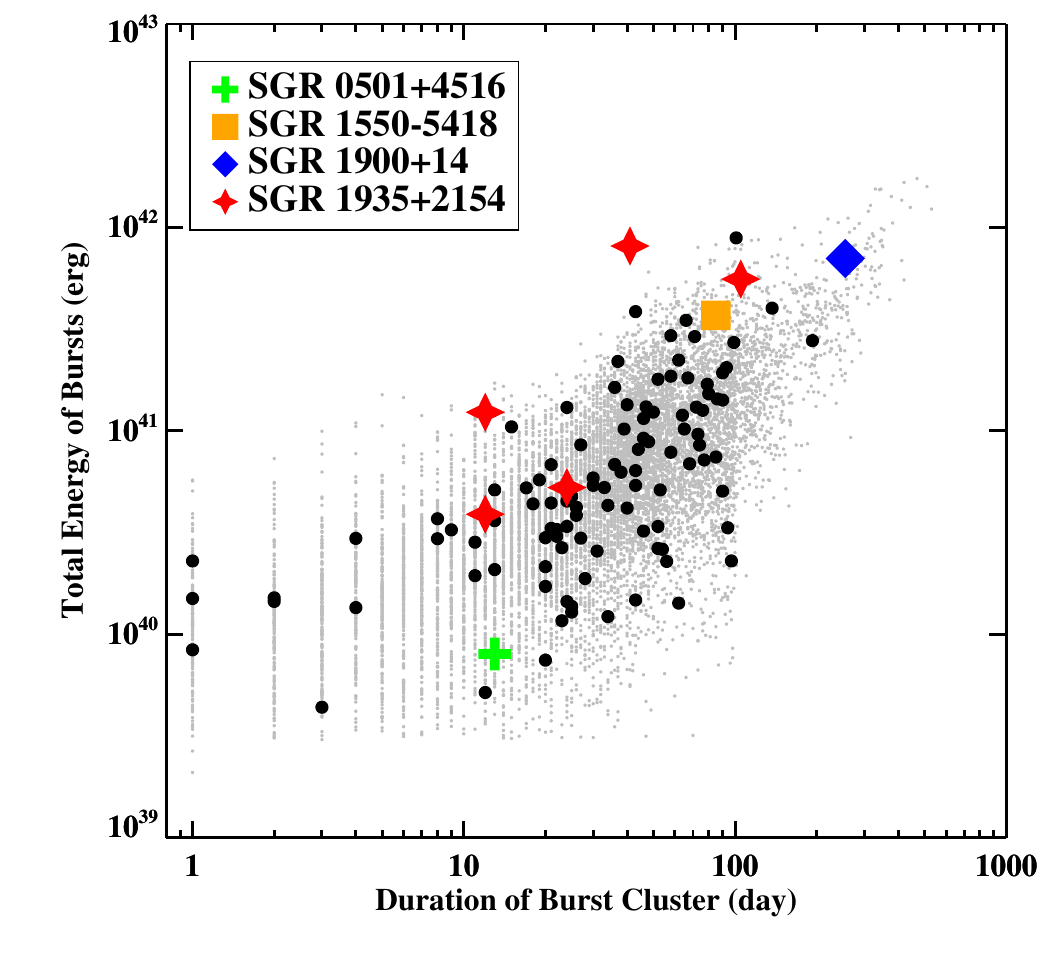}
    
    \caption{The total burst energy vs. duration of each burst cluster from 100 simulations through 1000-yr evolution (grey dots). The black circles show the results of a single simulation, the ones shown in Figure \ref{fig:burst_vs_time_2}. Colored data points represent the same relation for the actual observations of magnetars. [LEFT] Initial stress level: 0.7$\tau_{el}$ - 1.1$\tau_{el}$. [MIDDLE] Initial stress level: 0.5$\tau_{el}$ - 1.1$\tau_{el}$. [RIGHT] Another cell rule (see text for details).}
    \label{fig:burst_clus_2}
\end{figure}

We performed more simulations with different initial stress levels and cell rules to test whether we would always obtain a similar clustered bursting behavior of short bursts. During the 1000-yr evolution of a magnetar, our simulations recorded $\sim$ 12,500 bursts; in only 1000 of them, cells fail and enter a plastic phase alone (no plastic neighbor) while most have at least 1 plastic neighbor (simultaneous plastic phases). Therefore, bursts are prone to occur successively. This accumulation of bursts seems clearer (with a narrower time window around the years 40, 80, and 150; see Figure \ref{fig:burst_vs_time}) when the magnetar is young. This is due to a highly stressed crust at the beginning of the simulation: $\tau$ values of the cells are between 0.9$\tau_{el}$ - 1.1$\tau_{el}$. If we randomly set the initial stress level between 0.7$\tau_{el}$ - 1.1$\tau_{el}$, the simulation results in a similar total number of bursts, cell interactions, and the number of successive failures, compared to the original case. The difference is that we do not observe excessive burst accumulation when the magnetar is young (see the top panel of Figure \ref{fig:burst_vs_time_2}). As can be seen in the figure, we still observe burst clustering behavior; the only difference is that the number of bursts in a burst cluster when the magnetar is young (years $<$ 200) is similar to the burst clusters observed in later years. Therefore, burst clusters do not have energies above $10^{42}$ erg in the corresponding burst cluster energy vs. duration relation in the left panel of Figure \ref{fig:burst_clus_2}. If we randomly set the initial $\tau$ values within an even larger range between 0.5$\tau_{el}$ - 1.1$\tau_{el}$, then the number of bursts decreases to $\sim$ 10,000, with fewer bursts observed when the magnetar is young compared to older age due to cells trying to reach their yield stress during their long Hall phases. Yet, we still observe burst clusters (see the middle panels of Figures \ref{fig:burst_vs_time_2} and \ref{fig:burst_clus_2}). 

We also tried another cell rule as follows: If a cell has two or more (3 or 4) plastic neighbors, it fails above $\tau_{el}$; if it has one plastic neighbor, it fails at a $\tau$ value of 1.05$\tau_{el}$; otherwise, it fails at $\tau$ = 1.1$\tau_{el}$. This is effectively a slightly less nuanced version of the rule from \cite{Lander23}. In this case, the number of bursts is still $\sim$12,500, but the number of interactions decreased; the number of cells entering a plastic phase alone (without a plastic neighbor) is doubled. Again, we still observe burst clustering behavior, albeit less obvious in this case (see the bottom panel of Figure \ref{fig:burst_vs_time_2}). The right panel of Figure \ref{fig:burst_clus_2} represents the corresponding cluster energy vs. duration obtained with this cell rule.


\bibliographystyle{aasjournal}
\bibliography{refs} 

\end{document}